\pdfoutput=1

\documentclass[sigconf]{acmart}

\usepackage{booktabs} 
\usepackage{color}
\usepackage{soul}
\usepackage[skip=5pt]{caption}
\usepackage{adjustbox}
\usepackage{multirow}
\usepackage{graphicx}
\usepackage{subcaption}
\usepackage{tikz}
\usetikzlibrary{arrows,positioning,shapes.geometric}

\setcopyright{rightsretained}

\acmConference[]{Under Review}{2019}{}
\copyrightyear{2019}

\begin{document}

\title[Personalized QAC Through a Lightweight Representation of the User Context]{Personalized Query Auto-Completion Through a Lightweight Representation of the User Context}

\author{Manojkumar Rangasamy Kannadasan}
\affiliation{%
  \institution{eBay Inc.}
  \streetaddress{2025 Hamilton Avenue}
  \city{San Jose}
  \state{CA}
  \postcode{95125}
}
\email{mkannadasan@ebay.com}

\author{Grigor Aslanyan}
\affiliation{%
  \institution{eBay Inc.}
  \streetaddress{2025 Hamilton Avenue}
  \city{San Jose}
  \state{CA}
  \postcode{95125}
}
\email{gaslanyan@ebay.com}

\begin{abstract}
	Query Auto-Completion (QAC) is a widely used feature in many domains, including web and eCommerce search. This feature suggests full queries based on a prefix of a few characters typed by the user. QAC has been extensively studied in the literature in the recent years, and it has been consistently shown that adding personalization features can significantly improve the performance of the QAC model. In this work we propose a novel method for personalized QAC that uses lightweight embeddings learnt through \emph{fastText} \cite{bojanowski2017enriching,joulin2016bag}. We construct an embedding for the user context queries, which are the last few queries issued by the user. We also use the same model to get the embedding for the candidate queries to be ranked. We introduce ranking features that compute the distance between the candidate queries and the context queries in the embedding space. These features are then combined with other commonly used QAC ranking features to learn a ranking model using the state of the art LambdaMART ranker \cite{burges2010ranknet}. We apply our method to a large eCommerce search engine (eBay) and show that the ranker with our proposed feature significantly outperforms the baselines on all of the offline metrics measured, which includes Mean Reciprocal Rank (MRR), Success Rate (SR), Mean Average Precision (MAP), and Normalized Discounted Cumulative Gain (NDCG). Our baselines include the Most Popular Completion (MPC) model which is a commonly used baseline in the QAC literature, as well as a ranking model without our proposed features. The ranking model with the proposed features results in a $20-30\%$ improvement over the MPC model on all metrics. We obtain up to a $5\%$ improvement over the baseline ranking model for all the sessions, which goes up to about $10\%$ when we restrict to sessions that contain the user context. Moreover, our proposed features also significantly outperform text based personalization features studied in the literature before, and adding text based features on top of our proposed embedding based features results only in minor improvements.
\end{abstract}

\maketitle

\section{Introduction}\label{sec-intro}

Query Auto-Completion (QAC) is a common feature of most modern search engines. It refers to the task of suggesting full queries after the user has typed a prefix of a few characters \cite{INR-055}. QAC can significantly reduce the number of characters typed \cite{Zhang:2015:AAQ:2766462.2767697}, which is especially helpful to users on mobile devices. QAC can also help reduce the number of spelling errors in queries. In cases when the user is not really sure how to formulate the query, QAC can be of great help. It has been shown that QAC can greatly improve user satisfaction \cite{Song:2011:PQS:2009916.2010025}. Moreover, this can reduce the overall search duration, resulting in a lower load on the search engine \cite{Bar-Yossef:2011:CQA:1963405.1963424}. Currently QAC has a wide range of applications, including search (such as web, eCommerce, email), databases, operating systems, development environments \cite{INR-055}.

Query Auto-Completion has been extensively studied in the literature in the recent years. A detailed survey of the work prior to 2016 can be found in \cite{INR-055}, which broadly classifies QAC approaches into two main categories - heuristic models and learning based models. Heuristic models use a few different sources for each possible query completion and compute a final score. These approaches do not use a large variety of features. In contrast, learning based approaches treat the problem as a ranking problem and rely on the extensive research in the literature in the learning-to-rank (LTR) field \cite{INR-016}. Learning based approaches rely on a large number of features and generally outperform heuristic models \cite{INR-055}. The features for both of these approaches can be broadly characterized into three groups - time-sensitive, context-based, and demography based. Time-sensitive features model the query popularity and changes over time, such as weekly patterns. Demographic based features, such as gender and age, are typically limited and may be hard to access. In contrast, context based features rely on the user's previous search activity (short term, as well as long term) to suggest new query completions. This data is essentially free, making context-based features an attractive approach for personalizing QAC. Context-based features for LTR models will be the focus of this work.

In this paper we propose a novel method to learn the query embeddings \cite{bojanowski2017enriching,joulin2016bag} using a simple and scalable technique and use it to measure similarity between user context queries and candidate queries to personalize QAC. We learn the embeddings in a semi-supervised fashion using \emph{fastText} by taking all the queries in a session as a single document. We design features that measure the similarity between the context and candidate queries, which are then incorporated into a learning-to-rank model. We use the state of the art LambdaMART model \cite{burges2010ranknet} for ranking candidate queries for QAC. Even though embedding based features have been studied for QAC in the literature before, as discussed in Section \ref{sec-related}, our work makes the following novel contributions:
\begin{itemize}
	\item A lightweight and scalable way to represent the user's context in the embedding space.
	\item Simple and robust ranking features based on such embeddings for QAC, which can be used in any heuristic or LTR model.
	\item Training and evaluation of a pairwise LambdaMART ranker for QAC using the proposed features. We show that our proposed features result in significant improvements in offline metrics compared with state-of-the-art baselines.
	\item We also compare and combine text based features with embedding based features and show that embedding based features result in larger improvements in offline metrics.
\end{itemize}

The rest of the paper is organized as follows. Section \ref{sec-related} discusses some of the related work in the literature. In Section \ref{section-personalizedqac} we describe our methodology. In Section \ref{sec-experiments} we describe our datasets and experiments. We summarize our work and discuss possible future research in Section \ref{sec-summary}.

\section{Related Work}\label{sec-related}

The user's previously entered text is used for personalized QAC by Bar-Yossef and Kraus \cite{Bar-Yossef:2011:CQA:1963405.1963424}. The method, called NearestCompletion, computes the similarity of query completion candidates to the context queries (user's previously entered queries), using term-weighted vectors for queries and contexts and applying cosine similarity. This method results in significant improvements in MRR. In addition, the authors proposed the MPC approach, which is based on the overall popularity of the queries matching the given prefix. MPC is a straightforward heuristic approach with good performance and is typically used as a baseline for more complex approaches. We use MPC as one of the baselines in this work as well.

The user's long term search history is used in \cite{Cai:2016:SPQ:2911451.2914686} to selectively personalize QAC, where a trade-off between query popularity and search context is used to encode the ranking signal. Jiang et. al. \cite{Jiang:2014:LUR:2600428.2609614} study user reformulation behavior using textual features. Shokouhi \cite{Shokouhi:2013:LPQ:2484028.2484076} studies QAC personalization using a combination of context based textual features and demographic features, and shows that the user's long term search history and location are the most effective for QAC personalization. Su et. al. \cite{7363869} propose the framework EXOS for personalizing QAC, which also relies on textual features (token level). Jiang et. al. \cite{jiang-fei-honghui-2018} use history-level, session-level, and query-level textual features for personalized QAC. Fei et. al. \cite{cai2017learning} study features on the observed and predicted search popularity both for longer and shorter time periods for learning personalized QAC. Diversification of personalized query suggestion is studied in \cite{chen2017personalized}.

Recurrent Neural Networks (RNN) \cite{Jain:1999:RNN:553011} have also been studied for QAC. Three RNN models - session-based, personalized, and attention based, have been proposed in \cite{8577694}. Fiorini and Lu \cite{fiorini2018personalized} use user history based features as well as time features as input to an RNN model. \cite{park2017neural} uses RNNs to specifically improve QAC performance on previously unseen queries. An adaptable language model is proposed in \cite{jaech2018personalized} for personalized QAC.

Word embeddings, such as \emph{word2vec} \cite{Mikolov:2013:DRW:2999792.2999959}, \emph{glove} \cite{pennington2014glove}, and \emph{fastText} \cite{bojanowski2017enriching,joulin2016bag}, have become increasingly popular in the recent years for a large variety of tasks, including computing similarity between words. Embeddings have also been studied in the context of QAC. Specifically, Mitra \cite{mitra2015exploring} studies a Convolutional Latent Semantic Model for distributed representations of queries. Query similarity based on \emph{word2vec} embeddings is studied in \cite{shao2018query} where the features are combined with the MPC model. In Section \ref{section-personalizedqac} , we explain our approach of learning embeddings for the user context in a simple and scalable fashion and the usage of these embeddings and text based features to personalize QAC.

\section{Personalized Query Auto-Completion with Reformulation}\label{section-personalizedqac}

A search session is defined as a sequence of queries $\langle q_1, q_2, \dots, q_T\rangle$ issued by a user within a particular time frame. A query consists of a set of tokens. If the user types a prefix $p_T$ and ends up issuing the query $q_T$, then the user's context is the previous queries issued till time step $T$, $\langle q_1, q_2, \dots, q_{T-1}\rangle$. For example, if the queries issued in a session is $\langle nike, adidas, shoes\rangle$, the prefix used to issue the query $shoes$ is $sh$, then $\langle q_1, q_2, \dots, q_{T-1}\rangle = \langle nike, adidas\rangle$, $p_T = sh$, $q_T = shoes$. Given a prefix $p_T$, the user context $\langle q_1, q_2, \dots, q_{T-1}\rangle$ and candidate queries $Q_T$ matching the prefix, our goal is to provide ranking for the queries $q \in Q_T$ such that we have the best ranking for $q_T$. The ranking score can be considered as $P(Q_T | \langle q_1, q_2, \dots, q_{T-1}\rangle)$. This can be solved using the learning to rank framework. 

The influence of user context features towards the prediction accuracy has already been studied in \cite{Jiang:2014:LUR:2600428.2609614,mitra2015exploring}. In this paper we propose a simple and scalable way to understand the user's context using query embeddings and use multiple distance related features to compare the user's context to the candidate queries $Q_T$.

\subsection{Learning Query Representation for Reformulation}\label{subsection-learningqueryrep}

Continuous text representations and embeddings for a text can be learnt through both supervised \cite{mitra2015exploring} and semi-supervised approaches \cite{Mikolov:2013:DRW:2999792.2999959,pennington2014glove,bojanowski2017enriching,joulin2016bag}. In this paper, we learn the query representations via semi-supervised techniques. We use the publicly available \emph{fastText} library \cite{bojanowski2017enriching,joulin2016bag} for efficient representation learning to learn the query embeddings. The \emph{fastText} model learns subword
representations while taking into account morphology. The model considers subword
units, and represents a word by the sum of its character $n$-grams. The word $iphone$ with character $n$-grams $(n = 3)$ is represented as:
\begin{center}
 ``$\langle$ip'', ``iph'', ``pho'', ``hon'', ``one'', ``ne$\rangle$''
\end{center} 

 Some of the previous work learns distinct vector representations for the words thereby ignoring internal structure of the words \cite{Mikolov:2013:DRW:2999792.2999959}. If we have a dictionary of $n$-grams of size $G$, then the set of $n$-grams in a word $w$ is denoted as $Gr_w \in \{1, 2, \dots, G\}$. We use the \emph{skipGram} model where the goal is to independently predict the presence or absence of the context words. The problem is framed as a binary classification task. For the word at position $t$ we consider all context words as positive examples and sample negatives at random from the dictionary as described in \cite{bojanowski2017enriching,joulin2016bag}. For a context word $w_c$, we use the negative log likelihood, $l: x  \mapsto log(1+e^{-x})$, for the binary logistic loss. The objective function is defined as:
\begin{equation}
  \sum_{t=1}^{T}\left[ \sum_{c \in Context} l(s(w_t, w_c)) + \sum_{n \in N_{t,c}} l(-s(w_t, n))\right]
\end{equation}
where $w_t$ is the target word, $w_c$ is the context word, $N_{t,c}$ is a set of negative examples sampled from the vocabulary. The scoring function, $s(w, c)$ is defined as 
\begin{equation}
  s(w, c) = \sum_{g \in G_w} z_g^T v_c
\end{equation}
where $z_g$ is the vector representation of each $n$-gram of a word $w$ and $v_c$ is the vector representation of the context. Our goal is to learn scalable and lightweight embeddings for queries based on their reformulation behavior across different users. Here we represent all the queries issued in a session $\langle q_1, q_2, \dots, q_T\rangle$ as one document in the training data. By learning the subword representations using the probability of words in the context of other words present in the queries issued in the same context, we are able to provide a simple and scalable way to encode the query reformulation behavior in the embedding space. We are also able to learn the syntactic and semantic similarity between the vocabulary.

We learn query representations by mining 3 days of eBay's search logs to get the query reformulations issued by the user.  The query log is segmented into sessions with a 30 minute session boundary as followed in \cite{Jiang:2014:LUR:2600428.2609614}. Based on this definition of a session boundary, we collect different queries issued by the users within that session. We remove special characters in the query and convert them to lowercase. We also filter out sessions with only one query in the session. For example, if the user issues $q_1, q_2, \dots, q_T$ in a session, then all of these queries $\langle q_1, q_2, \dots, q_T\rangle$ together, separated by whitespace, are considered as one user context. For example, if a session contains 2 queries in the user context, ``\emph{iphone}'', ``\emph{iphone xs case}'', then a single document for training will be represented as ``\emph{iphone iphone xs case}''.

For training unsupervised character \emph{$n$-Gram} representations we consider each user context as one document sample. We tune the model hyperparameters by fixing the dimension of subword representations as 50, minimum occurrence of the words in the vocabulary to be 20 and hierarchical softmax as the loss function. The other hyperparameters of the model are tuned based on the \textit{Embedding\_Features} model described in Section \ref{subsection-personalizedranker}. The number of unique words in the vocabulary used to train the model is 189,138. The user context $\langle q_1, q_2, \dots, q_T\rangle$ is then converted to multiple vector representations. Similar vector representations are also created for all candidate target queries in the dataset.

\begin{table}
	\begin{center}
		\caption{Textual features based on the user context defined across 3 categories. A subset of features are highlighted in the table. Rest of the features are derived from them. }
		\label{dim-table}
		\resizebox{\columnwidth}{!}{
		\begin{tabular}{|c|c|}
			\hline
			Category & Examples \\
			\hline
			\begin{tabular}{@{}c@{}} Token \\ (16 features) \end{tabular} &  \begin{tabular}{@{}c@{}} ratio of new terms \\ ratio of used terms \\ average terms in previous queries \\ median terms in previous queries \\ trend of number of terms \\ unique terms added from last query \\ unique terms retained from last query \\ unique terms removed from last query \\ unique terms added from all previous queries \\ occurrence of terms in previous queries  \end{tabular}  \\
			\hline
			\begin{tabular}{@{}c@{}} Query \\ (7 features) \end{tabular} & \begin{tabular}{@{}c@{}} frequency in previous queries \\ character n-gram similarity with previous queries \\ token n-gram similarity with previous queries  \end{tabular} \\
			\hline
			\begin{tabular}{@{}c@{}} Session \\ (3 features) \end{tabular} & \begin{tabular}{@{}c@{}} position in session \\ unique terms in session \\ common terms in session \end{tabular}  \\
			\hline
		\end{tabular}}
	\end{center}
	\vspace{-5mm}
\end{table}

\subsection{User Context Features}\label{subsection-usercontextrep}

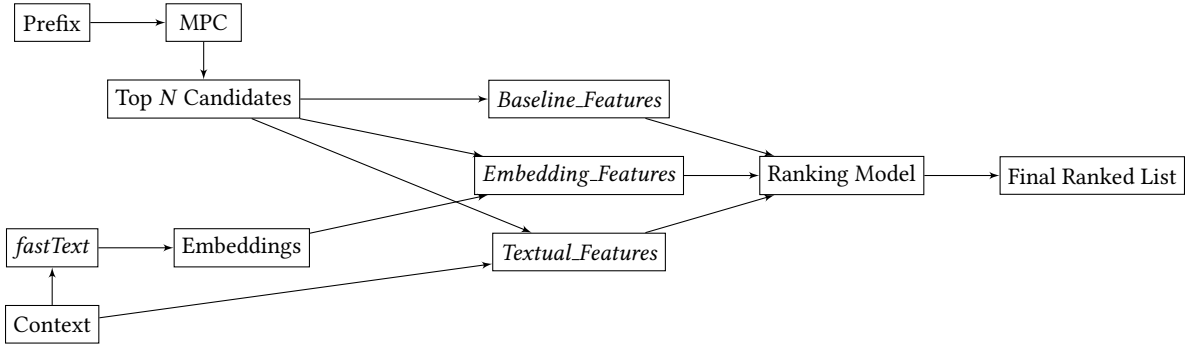
\begin{figure*}
\begin{tikzpicture}[>=latex']
        \tikzset{block/.style= {draw, rectangle, align=center,minimum width=1cm,minimum height=0.5cm}}
        \node [block]  (prefix) {Prefix};
	\node [block, right = 1.0cm of prefix]  (mpc) {MPC};
	\node [block, below = 0.5cm of mpc]  (topn) {Top $N$ Candidates};
	\node [block, right = 2.5cm of topn]  (baselinefeat) {\emph{Baseline\_Features}};
	\node [block, below = 0.5cm of baselinefeat]  (embfeat) {\emph{Embedding\_Features}};
	\node [block, below = 0.5cm of embfeat]  (textfeat) {\emph{Textual\_Features}};
	\node [block, below = 3.5cm of prefix]  (context) {Context};
	\node [block, above = 0.5cm of context]  (fasttext) {\emph{fastText}};
	\node [block, right = 1.0cm of fasttext]  (embeddings) {Embeddings};
	\node [block, right = 1.0cm of embfeat]  (model) {Ranking Model};
	\node [block, right = 1.0cm of model]  (final) {Final Ranked List};

	\path[draw, ->] (prefix) -- (mpc);
	\path[draw, ->] (mpc) -- (topn);
	\path[draw, ->] (context) -- (fasttext);
	\path[draw, ->] (fasttext) -- (embeddings);
	\path[draw, ->] (topn) -- (baselinefeat);
	\path[draw, ->] (topn) -- (embfeat);
	\path[draw, ->] (topn) -- (textfeat);
	\path[draw, ->] (embeddings) -- (embfeat);
	\path[draw, ->] (context) -- (textfeat);
	\path[draw, ->] (baselinefeat) -- (model);
	\path[draw, ->] (embfeat) -- (model);
	\path[draw, ->] (textfeat) -- (model);
	\path[draw, ->] (model) -- (final);

\end{tikzpicture}
\caption{The end to end architecture of the \textbf{\emph{Textual\_Embedding}} model. The architecture for the other models is similar, except that some of the features will not be excluded.}\label{fig-architecture}
\end{figure*}

In this section we propose different user context features based on the queries issued in the session. Vector representations are created for both the individual queries as well as the entire context taking all queries in the session. $v_C$ represents the user context vector and $v_{q_T}$ represents the vector for one query at time step $T$. We develop four features based on the query representations learned in the previous section. We denote these features as \emph{Embedding Features}. One embedding feature is based on all the queries in the context. Since the median number of searches in a session is approx 3, we considered up to 3 queries previously issued by the user for generating the remaining embedding features. The \emph{Embedding Features} are computed as a distance between 2 vectors using \emph{cosine similarity} \cite{singhal2001modern}.

\begin{itemize}
\item \textbf{\emph{user\_context\_cosine}}: Cosine distance between the user context vector $v_C$ and the current target query $v_{q_T}$.

\item \textbf{\emph{prev\_query1\_cosine}}: Cosine distance between the query vector $v_{q_{T-1}}$ and the current target query $v_{q_T}$.

\item \textbf{\emph{prev\_query2\_cosine}}: Cosine distance between the query vector $v_{q_{T-2}}$ and the current target query $v_{q_T}$.

\item \textbf{\emph{prev\_query3\_cosine}}: Cosine distance between the query vector $v_{q_{T-3}}$ and the current target query $v_{q_T}$.

\end{itemize}

In addition to the \textbf{\emph{Embedding\_Features}}, we also developed various \textbf{\emph{Textual\_Features}} comparing the user context and the current target query to be ranked as defined in Table \ref{dim-table}. We categorize them into three categories, namely \emph{Token}, \emph{Query}, and \emph{Session}. There is a large overlap between the features defined in Table \ref{dim-table} and the features defined in \cite{Jiang:2014:LUR:2600428.2609614,Shokouhi:2013:LPQ:2484028.2484076}. A query $q_T$ can contain multiple tokens. Users may add or remove tokens between 2 consecutive queries in a session. Based on analyzing the user sessions, between queries $q_T$ and $q_{T-1}$, tokens can either be added and/or removed. These token reformulation user behavior can be encoded via 16 features, described in Table \ref{dim-table}, representing the effectiveness of the tokens in the context $C$ and the target query $q_T$. Similarly, Query level features represent how users reformulate the queries in a session through repetition and textual similarity between $q_T$ and $q_{T-1}$. The Session level features represent how users reformulate their queries without taking into account the relationship to the target query $q_T$.

\section{Experiments}\label{sec-experiments}
\subsection{Dataset and Experiment Setting}\label{subsection-datasetexpsetting}

We conduct our ranking experiments on a large scale search query dataset sampled from the logs of eBay Search engine. The query log is segmented into sessions with a 30 minute session boundary as described in \cite{Jiang:2014:LUR:2600428.2609614}. For ranking experiments, we do not filter out sessions containing a single query. This is to make sure that we have a single learning to rank model powering sessions with and without user context. The dataset obtained based on the above logic results in about $53\%$  of the sessions with user's context. This gives us good coverage of user context features to learn a global model.

The labeling strategy used in \cite{Jiang:2014:LUR:2600428.2609614,Shokouhi:2013:LPQ:2484028.2484076} assume there is at least one query in the context, remove target queries $q_T$ not matching the prefix $p_T$, setting the first character of the prefix $p_T$ based on $q_T$. In our method, we use a slightly different labeling strategy for building the personalized QAC. We sample a set of impressions from search logs. For each issued query $q_T$, we capture the prefix $p_T$ that led to the search page. This is marked as a positive label. For the same prefix $p_T$ we identify rest of the candidate queries $Q_T \setminus {q_T}$  that were shown to the user and did not lead to the impression. They are marked as negative labels. We also retain sessions without user context.

The above training data now consists of labeled \emph{prefix-query} pairs. To learn the performance of the lightweight query representation of reformulations, we use LambdaMART \cite{burges2010ranknet} as the choice of learning to-rank algorithms, a boosted tree version of LambdaRank. LambdaMART is considered as one of the state-of-the-art learning to rank algorithms and has won the Yahoo! Learning to Rank Challenge (2010) \cite{Jiang:2014:LUR:2600428.2609614}. We use a pairwise ranking model and fine tune our parameters based on the \textbf{\emph{Baseline\_Ranker}} defined in Section \ref{subsection-baseline}. We fix these parameters to train and evaluate our models across all of our experiments.

\subsection{Baseline System}\label{subsection-baseline}

To evaluate our new personalized QAC ranker we establish two baseline ranking algorithms.

\begin{itemize}
\item \textbf{\emph{MPC}}: The \emph{Most Popular Completion} model \cite{Bar-Yossef:2011:CQA:1963405.1963424} predicts and provides users with candidate queries which are ranked by the popularity of the query. Popularity of a query is defined as the number of times the query has been issued by all the users in the past.

\item \textbf{\emph{Baseline\_Ranker}}: The baseline ranker is a Learning to Rank model built using the same methodology for creating and labeling the dataset. The features used in building the model are prefix features, target query features and prefix-query features. We refer to these features as \emph{Baseline\_Features}. The hyperparameters used for the LambdaMART model are exactly the same as in all the experiments for the personalized ranker.
\end{itemize}

\subsection{Personalized Ranking Models}\label{subsection-personalizedranker}

We have developed three personalized ranking models with different combinations of user context features, as described in Section \ref{subsection-usercontextrep}. These ranking models are compared against the two baseline rankers by experimentally evaluating the improvements on eBay datasets. The results are presented in Section \ref{subsection-results}.

\begin{itemize}
\item \textbf{\emph{Textual}}: Ranker with \emph{Baseline\_Features} and \emph{Textual\_Features} representing the user context.

\item \textbf{\emph{Embedding}}: Ranker with \emph{Baseline\_Features}, as well as \emph{Embedding\_Features} representing the user context.

\item \textbf{\emph{Textual\_Embedding}}: Ranker with \emph{Baseline\_Features}, \emph{Textual\_Features}, and \emph{Embedding\_Features} representing the user context.
\end{itemize}

For all of the ranking models we first get the top $N$ candidate queries from the \textbf{\emph{MPC}} model and re-rank them with the ranking model. We show the full end to end architecture for the \textbf{\emph{Textual\_Embedding}} model in Figure \ref{fig-architecture}. The architecture for the other models is similar except that they will only include a subset of the features.

\subsection{Evaluation Metrics}\label{subsection-evalmetrics}

\begin{table*}[ht!]
\centering
\caption{Offline evaluation metrics. We show the ratio of the metrics for four of the ranking models to the MPC model on both test datasets - all data and filtered data to include full coverage for user context.}\label{table-results}
	\resizebox{\textwidth}{!}{
        \begin{tabular}{llllll}
            \hline
            Dataset & Measure  & Baseline\_Ranker & Textual\_Features & Embedding\_Features & Textual\_Embedding\_Features \\\hline
            \multirow{7}{*}{Whole} & \multicolumn{1}{l}{MRR} & \multicolumn{1}{l}{1.26} & \multicolumn{1}{l}{1.30} & \multicolumn{1}{l}{1.31} & \multicolumn{1}{l}{1.31} \\\cline{2-6}
				   & \multicolumn{1}{l}{nDCG} & \multicolumn{1}{l}{1.30} & \multicolumn{1}{l}{1.32} & \multicolumn{1}{l}{1.33} & \multicolumn{1}{l}{1.33} \\\cline{2-6}
                                 & \multicolumn{1}{l}{SR@1} & \multicolumn{1}{l}{1.24} & \multicolumn{1}{l}{1.29} & \multicolumn{1}{l}{1.31} & \multicolumn{1}{l}{1.31} \\\cline{2-6}
                                 & \multicolumn{1}{l}{SR@3} & \multicolumn{1}{l}{1.23} & \multicolumn{1}{l}{1.26} & \multicolumn{1}{l}{1.27} & \multicolumn{1}{l}{1.27} \\\cline{2-6}
                                 & \multicolumn{1}{l}{MAP} & \multicolumn{1}{l}{1.26} & \multicolumn{1}{l}{1.30} & \multicolumn{1}{l}{1.31} & \multicolumn{1}{l}{1.31} \\\cline{2-6}
                                 & \multicolumn{1}{l}{MAP@1} & \multicolumn{1}{l}{1.24} & \multicolumn{1}{l}{1.29} & \multicolumn{1}{l}{1.31} & \multicolumn{1}{l}{1.31} \\\cline{2-6}
                                 & \multicolumn{1}{l}{MAP@3} & \multicolumn{1}{l}{1.23} & \multicolumn{1}{l}{1.27} & \multicolumn{1}{l}{1.29} & \multicolumn{1}{l}{1.29} \\\hline
           \multirow{7}{*}{User Context Only} & \multicolumn{1}{l}{MRR} & \multicolumn{1}{l}{1.27} & \multicolumn{1}{l}{1.34} & \multicolumn{1}{l}{1.37} & \multicolumn{1}{l}{1.37} \\\cline{2-6}
                                 & \multicolumn{1}{l}{nDCG} & \multicolumn{1}{l}{1.30} & \multicolumn{1}{l}{1.35} & \multicolumn{1}{l}{1.37} & \multicolumn{1}{l}{1.37} \\\cline{2-6}
                                 & \multicolumn{1}{l}{SR@1} & \multicolumn{1}{l}{1.26} & \multicolumn{1}{l}{1.37} & \multicolumn{1}{l}{1.42} & \multicolumn{1}{l}{1.41} \\\cline{2-6}
                                 & \multicolumn{1}{l}{SR@3} & \multicolumn{1}{l}{1.23} & \multicolumn{1}{l}{1.30} & \multicolumn{1}{l}{1.33} & \multicolumn{1}{l}{1.34} \\\cline{2-6}
                                 & \multicolumn{1}{l}{MAP} & \multicolumn{1}{l}{1.27} & \multicolumn{1}{l}{1.34} & \multicolumn{1}{l}{1.37} & \multicolumn{1}{l}{1.37}  \\\cline{2-6}
                                 & \multicolumn{1}{l}{MAP@1} & \multicolumn{1}{l}{1.26} & \multicolumn{1}{l}{1.37} & \multicolumn{1}{l}{1.42} & \multicolumn{1}{l}{1.41} \\\cline{2-6}
                                 & \multicolumn{1}{l}{MAP@3} & \multicolumn{1}{l}{1.24} & \multicolumn{1}{l}{1.33} & \multicolumn{1}{l}{1.37} & \multicolumn{1}{l}{1.37} \\\hline
        \end{tabular}}
 \vspace{ - 4 mm}
\label{table:2}
\end{table*}

The quality of our predictions can be measured using the following metrics: 
\begin{itemize}
	\item \emph{Mean Reciprocal Rank (MRR)} - the average of the reciprocal ranks of the target queries in the QAC results. Given a test dataset $S$, the MRR for algorithm $A$ is computed as
\begin{equation}
  MRR(A) =\frac{1}{|S|}  \sum_{C_T, q_T \in S} \frac{1}{hitRank(A, C_T, q_T)}
\end{equation}
where $C_T$ represents the user context at time step $T$, $q_T$ represents the relevant target query, and the function $hitRank$ computes the rank of the relevant query based on the order created by algorithm $A$. Relevant query refers to the clicked query in QAC.
 
\item \emph{Success Rate at Top-K (SR@$K$)} - the average percentage of relevant queries ranked at or above the position $K$ in the ranked list from QAC. In this paper we will consider only SR@$1$, SR@$2$, SR@$3$.

\item \emph{Normalized Discounted Cumulative Gain ($nDCG$)} - The Discounted Cumulative Gain ($DCG$) represents the usefulness or gain of the query based on its position in the ranked list from QAC. $DCG$ penalizes the relevance of the query logarithmically based on the position of the query in the ranked list. $DCG$ is defined as 
\begin{equation}
  DCG_q = \sum_{i = 1}^{P} \frac{2^{rel_i}-1}{log_2(i+1)}
\end{equation} 
where $i$ denotes the rank and $rel_i$ is the relevance of query at rank $i$. For our purposes $rel_i$ takes values $0$ or $1$.
 
$nDCG$ is defined as normalized $DCG$. Namely, it is the ratio of $DCG$ to $IDCG$ (ideal $DCG$):
\begin{equation}
  nDCG_q = \frac{DCG_q}{IDCG_q}
\end{equation} 
where $IDCG_q$ is the maximum possible value of $DCG_q$ for any ranker.

The overall performance of the ranking algorithm $A$ is measured by the average $nDCG$ across all queries in the dataset:
\begin{equation}
  nDCG = \frac{\sum_{q = 1}^{Q} nDCG_q}{Q}\,.
\end{equation} 

\item \emph{Mean Average Precision (MAP)} - the mean of the average precision scores for each query across the entire set of queries:
\begin{equation}
  MAP = \frac{\sum_{q = 1}^{Q} AvgPrecision(q)}{Q}\,.
\end{equation} 

\end{itemize}

\subsection{Results}\label{subsection-results}

We perform our evaluation in two phases. Firstly, we evaluate the quality of our query representations. Secondly, we evaluate the user context embeddings against the user context based textual features using a Learning to Rank framework \cite{burges2010ranknet}. 

\begin{figure}
	\includegraphics[height=5.3cm, width=8.5cm]{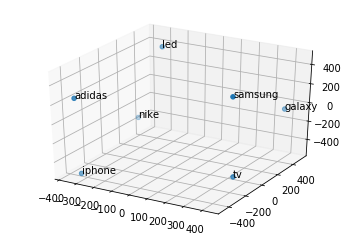}
	\setlength{\belowcaptionskip}{-12pt}
	\caption{A three-dimensional t-SNE plot using the vectors learned from user query reformulation, showing how similar intent words are modeled in the embedding space.}\label{fig-tsne}
\end{figure}

To evaluate our query representations we sample a few words across different verticals like fashion, electronics, home and garden, to evaluate if the embeddings are representing the syntactic and semantic knowledge of the queries learnt from the query reformulation behavior. We use t-SNE \cite{maaten2008visualizing} to visualize the embeddings for these sampled queries and show that words like \emph{samsung}, \emph{galaxy}, \emph{tv} are close to each other in the embedding space and far from queries like \emph{adidas} and \emph{iphone}. This verifies that our query embeddings have good subword information to represent the user context in the embedding space. The t-SNE plot for a small sample of queries is shown in Figure \ref{fig-tsne}.

\begin{figure*}
	\begin{tabular}{cc}
		\includegraphics[height=5cm, width=8cm]{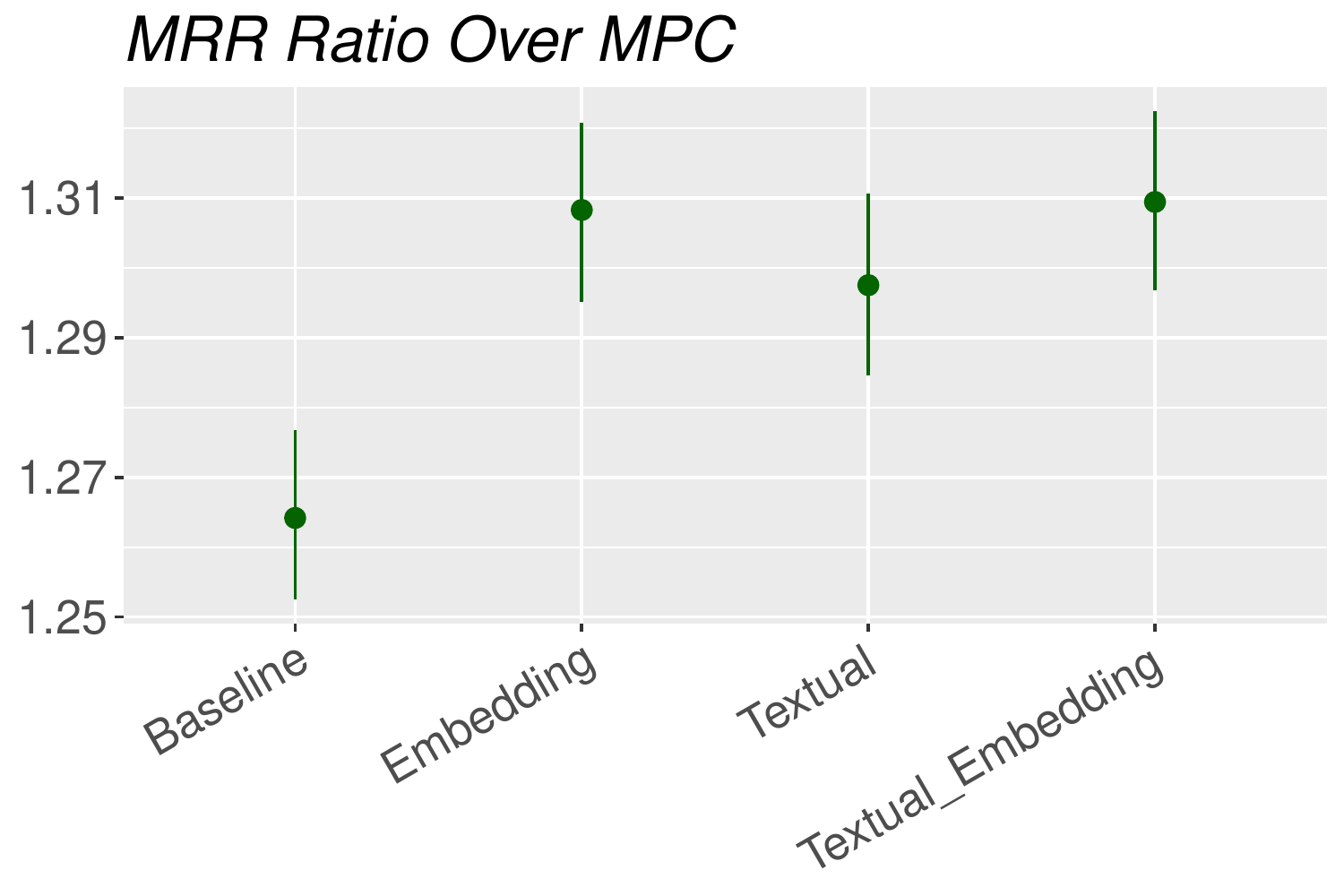} & \includegraphics[height=5cm, width=8cm]{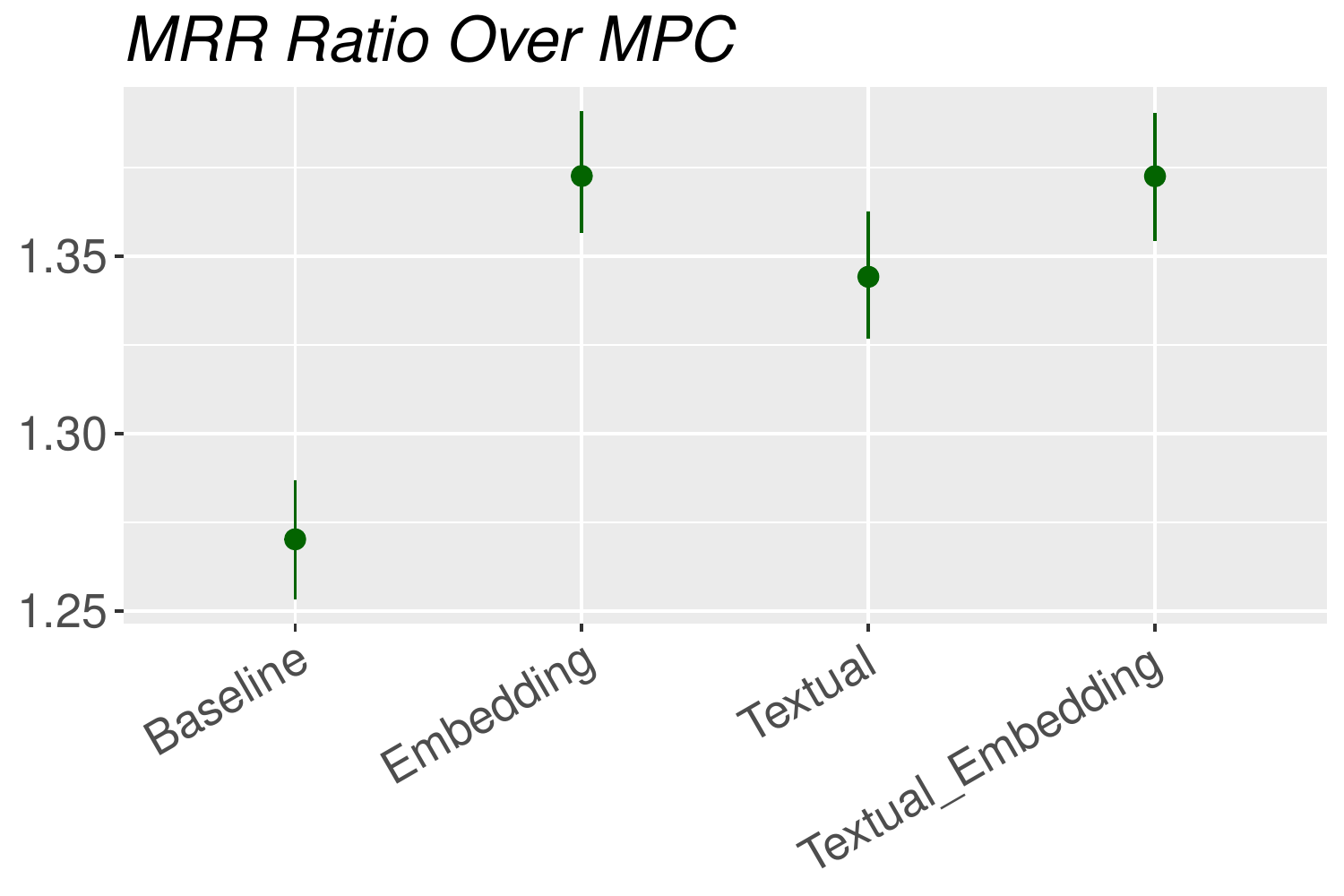} \\
		\includegraphics[height=5cm, width=8cm]{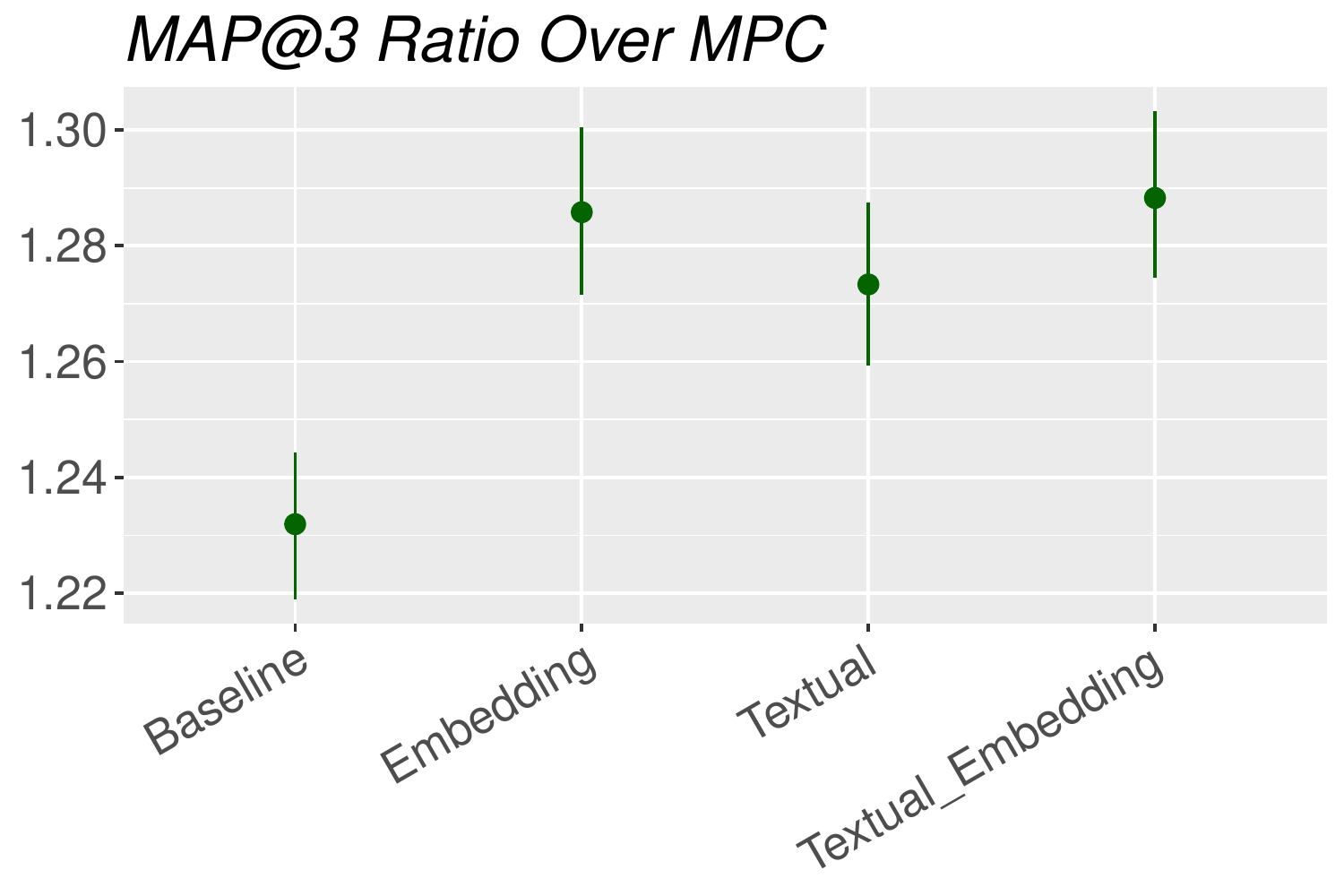} & \includegraphics[height=5cm, width=8cm]{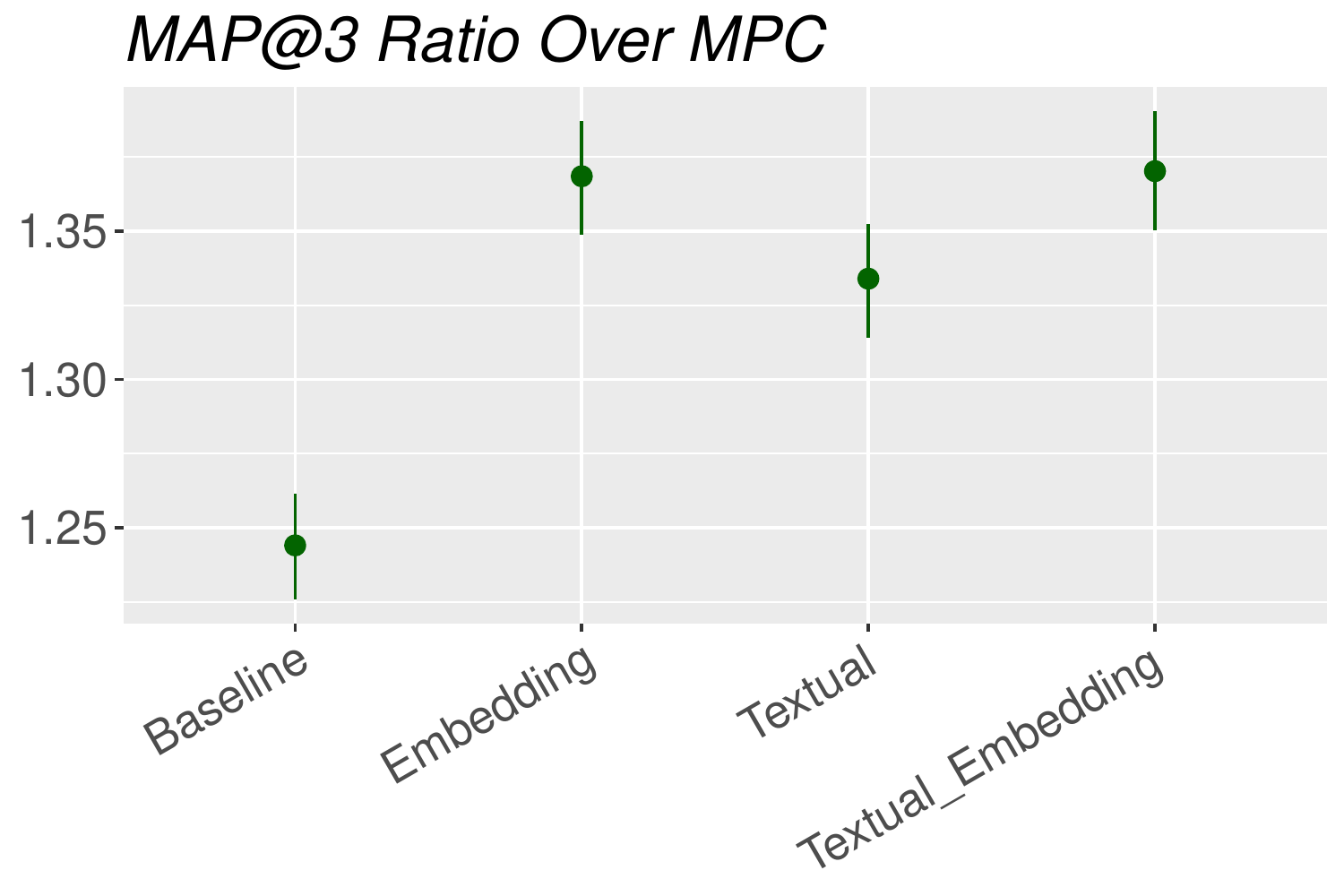} \\
		\includegraphics[height=5cm, width=8cm]{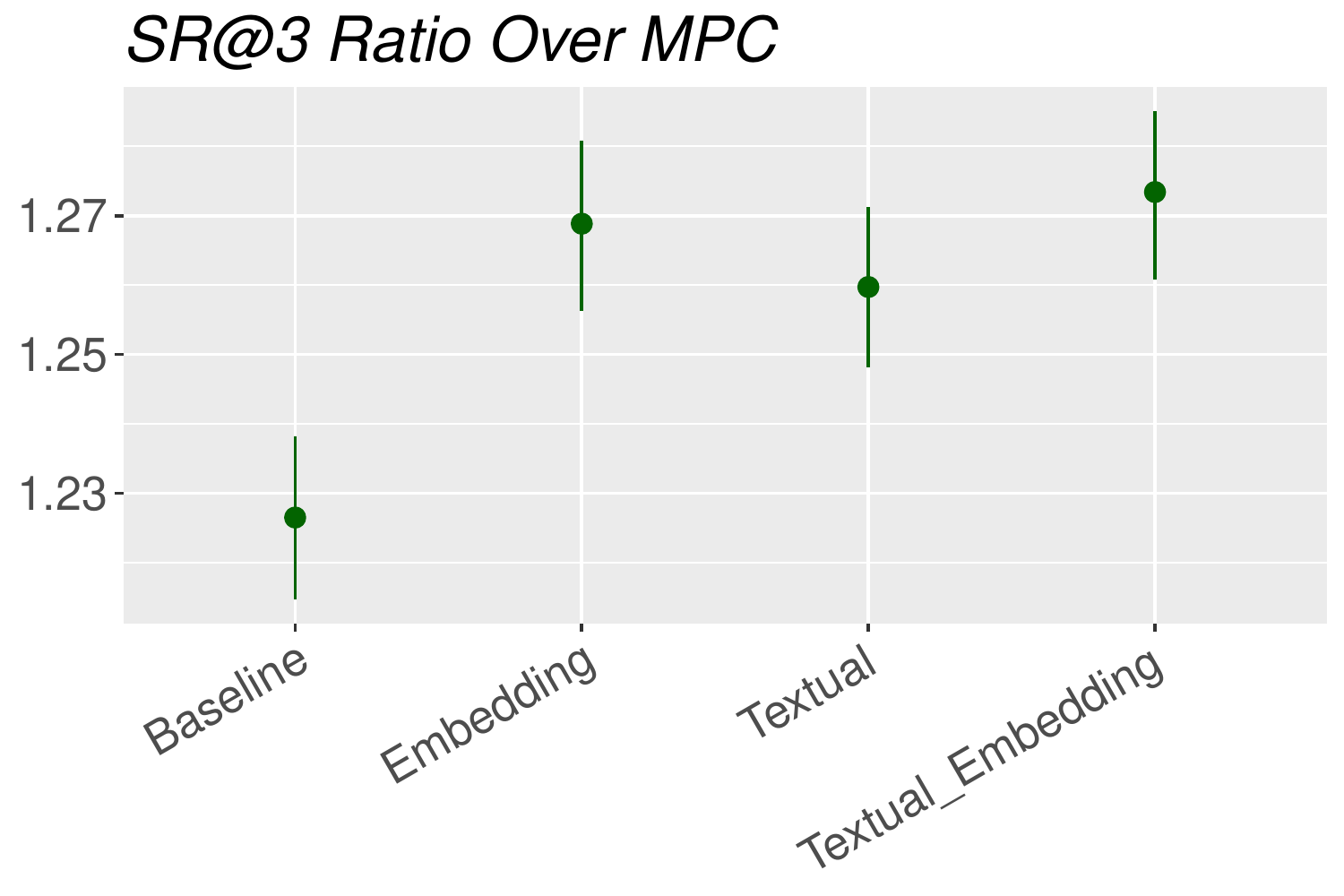} & \includegraphics[height=5cm, width=8cm]{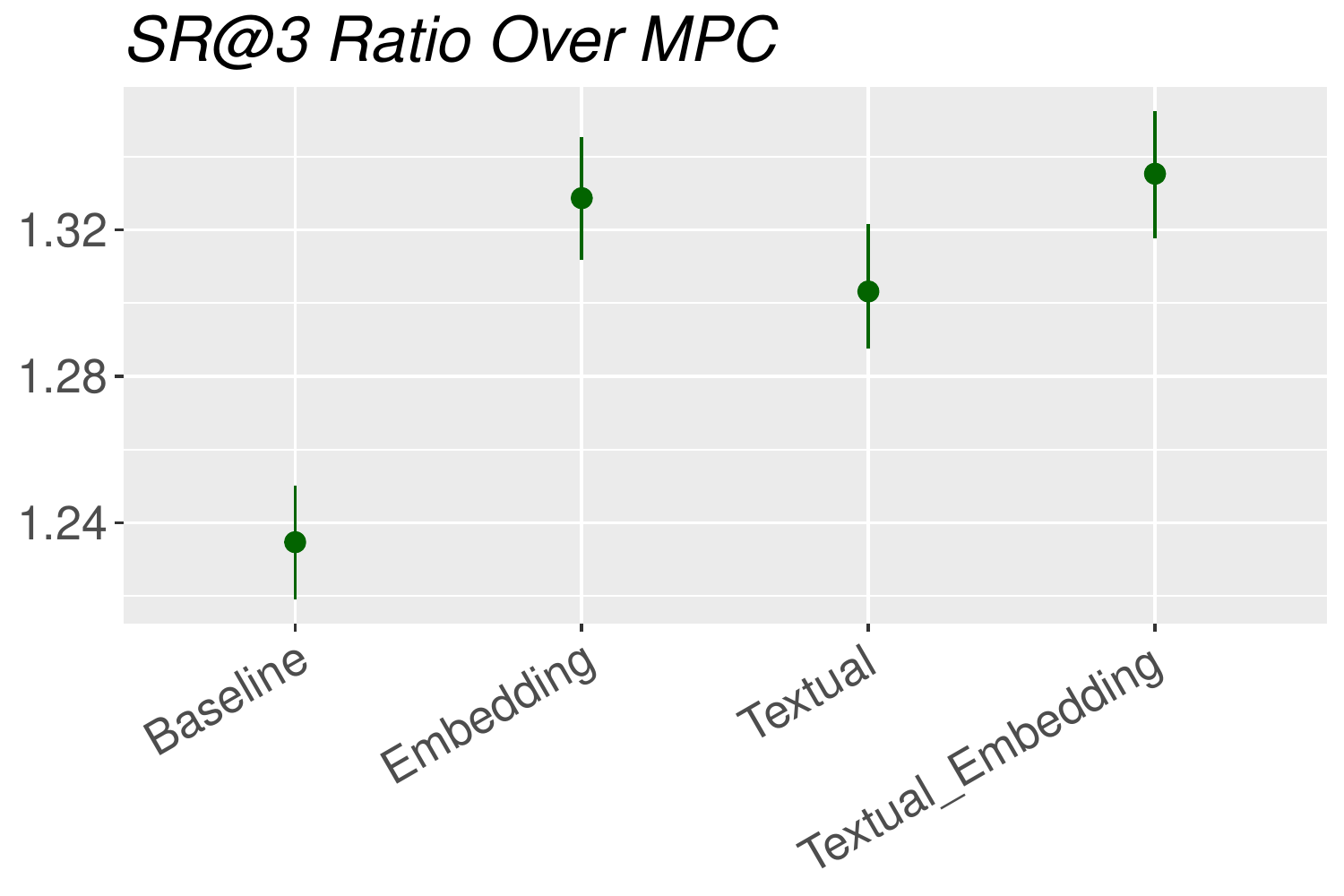} \\
		\includegraphics[height=5cm, width=8cm]{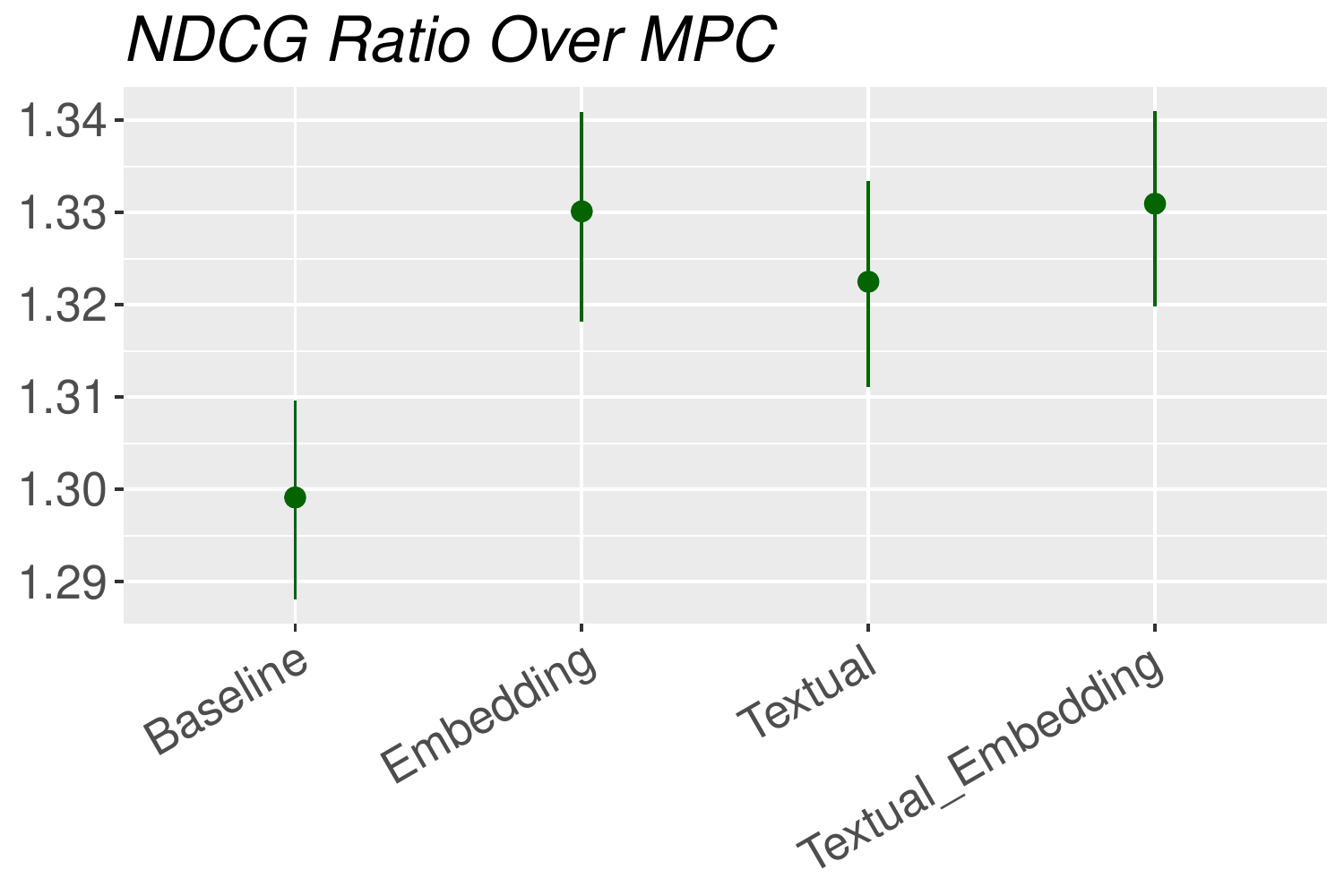} & \includegraphics[height=5cm, width=8cm]{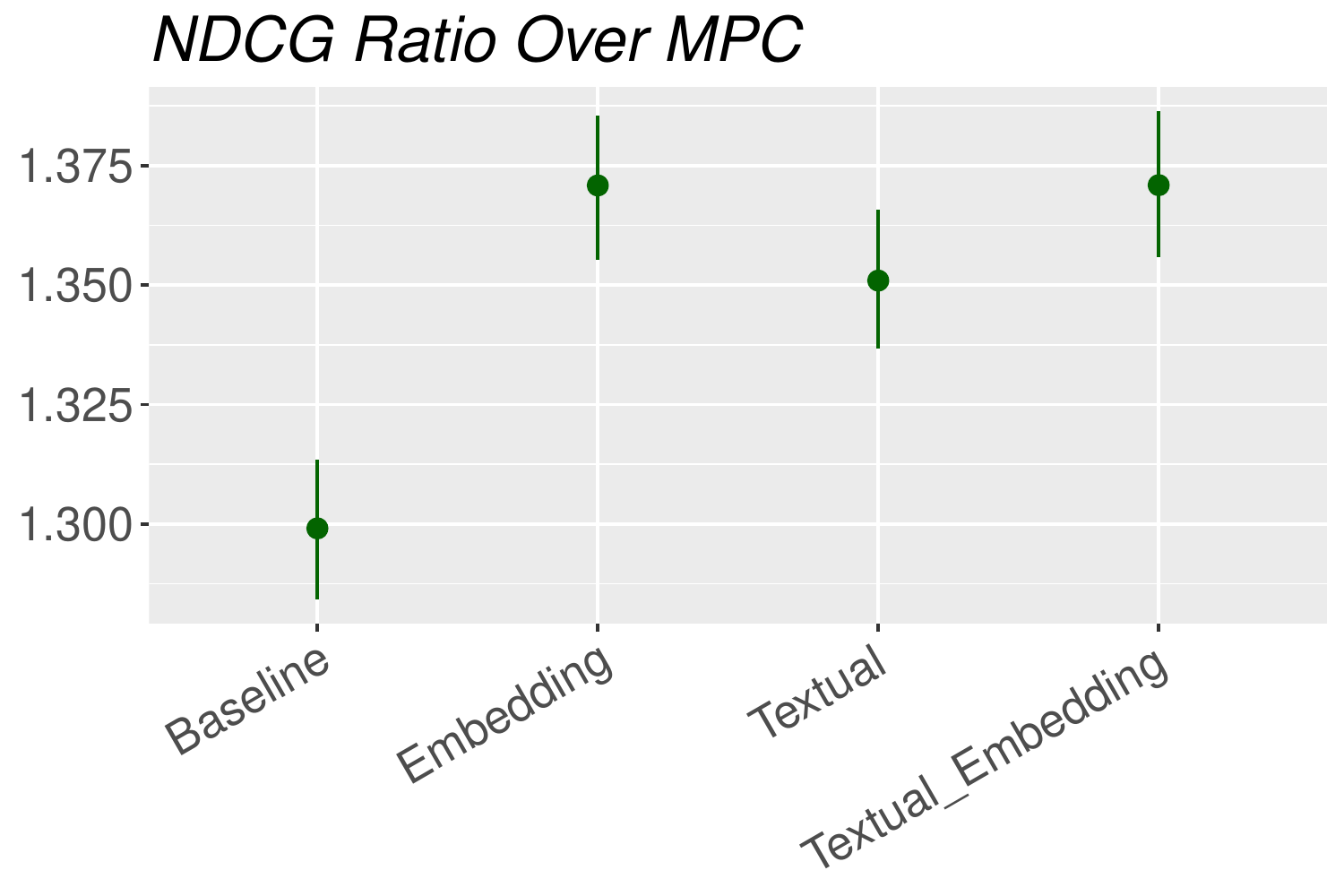}
	\end{tabular}
	\setlength{\belowcaptionskip}{-12pt}
	\caption{Metrics ratio to MPC for the whole dataset (left) and the user context only dataset (right). Error bars are computed using 1,000 bootstrap samples of the test queries.}\label{fig-metrics-whole}
\end{figure*}

Offline metrics $MRR$, $SR@k$, $nDCG$, $MAP$, and $MAP@k$ are shown in Table \ref{table-results}, where we have normalized the metrics with respect to the MPC model. We show results for the whole test dataset, which includes queries with and without user context, as well as the dataset with user context only. To assess statistical significance we have performed 1,000 bootstrap samples over the test queries and computed $95\%$ confidence intervals using those samples. The metrics, together with the $95\%$ confidence intervals, are plotted in Figure \ref{fig-metrics-whole}, where the plots on the left are for the whole dataset and the plots on the right are for the context only dataset. We have only plotted one variant of each metric since the others are very similar.

Our results show that all of the LTR models result in $20-30\%$ improvements over the MPC model. All three models with contextualization features outperform the \textbf{\emph{Baseline\_Ranker}} on all the metrics statistically significantly. For example, for $MAP@3$ \textbf{\emph{Embedding}} outperforms \textbf{\emph{Baseline\_Ranker}} by $5\%$ on the whole dataset and $10\%$ for the context only dataset. The \textbf{\emph{Embedding}} model also outperforms \textbf{\emph{Textual}} with an improvement of $1.5\%$ for $MAP@3$ on the whole dataset and $3\%$ for the context only dataset. The \textbf{\emph{Textual\_Embedding}} model performs very similarly to \textbf{\emph{Embedding}} which implies that the embedding based features proposed in this work capture all of the information in the textual features (from the perspective of the ranking model), and provide additional significant improvements.

\subsection{Feature Analysis}\label{subsection-featanal}

\begin{figure}[h]
\centering
\begin{tabular}{cc}
    \includegraphics[width=0.47\linewidth]{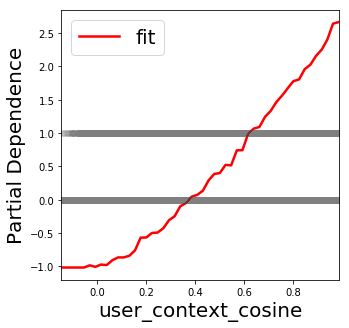}&
    \includegraphics[width=0.47\linewidth]{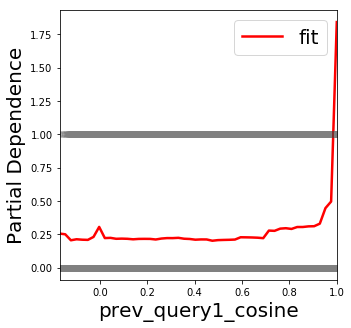}\\[2\tabcolsep]
    \includegraphics[width=0.47\linewidth]{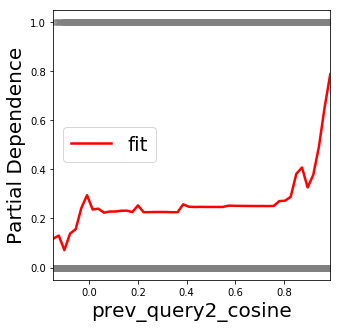}&
    \includegraphics[width=0.47\linewidth]{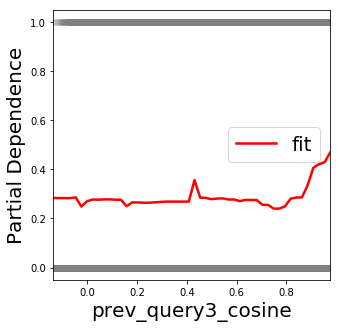}
\end{tabular}
\caption{Partial dependence plots for user context embedding features learnt from query reformulation.}\label{fig-pdp}
\end{figure}

In this section we analyze the user context embedding features through partial dependence plots shown in Figure \ref{fig-pdp}. The partial dependence plot for the \emph{user\_context\_cosine} feature clearly indicates that the cosine similarity between the user context $\langle q_1, q_2, \dots, q_{T-1}\rangle$ and the target query $q_T$ has a linear relationship with the target. The embedding features based on individual time step (\emph{prev\_query1\_cosine}, \emph{prev\_query2\_cosine}, \emph{prev\_query3\_cosine}) also show a clear monotonic relationship.

\section{Summary and Future Work}\label{sec-summary}

In this work we have presented a simple and scalable approach to learn lightweight vector representations (embeddings) for the query reformulations in a user session. These query representations exhibit both syntactic and semantic similarities between different queries and enable them to model the user context seamlessly. We have leveraged these lightweight embeddings to represent the user context in our personalized ranker for Query Auto-Completion. Different combinations of user context features are created, including textual and embedding features on top of our baseline ranker. We have applied these personalization features to a large scale commercial search engine (eBay) and experimentally verified significant improvements on all the offline ranking metrics. We have evaluated our personalized ranker on the entire dataset and a dataset restricted to sessions containing the user context. We see the biggest improvements on the user context filtered dataset. Furthermore, we show that the ranking model with embedding features outperforms the model with the textual features, whereas the model with combined textual and embedding features results in only minor improvements on top of the model with embedding features alone. The minor improvements from the textual features is likely due to the session level features which are agnostic of the queries in the context. As a future work, we would like to explore different representation learning techniques like \emph{sent2vec}, \emph{doc2vec}, and sequence models, to understand the user context better and incorporate them in the personalized ranker. We also plan to explore the trade offs between short term and long term user contexts in QAC. Lastly, the user context vectors provide a simple and scalable way to understand the user sessions which can be utilized for personalizing different parts of search and recommender systems.

\bibliographystyle{ACM-Reference-Format}
\bibliography{bibliography}

\end{document}